\RequirePackage{etex} 
\documentclass[runningheads,twocolumn,a4paper,10pt]{llncs}
\usepackage[maxnames=6,firstinits=true,doi=false,url=true,isbn=false]{biblatex}
\usepackage{etex}
\reserveinserts{28}
\pdfoutput=1
\usepackage[a4paper, total={7in, 10in}]{geometry}
\usepackage[linesnumbered,lined,boxed,commentsnumbered,ruled]{algorithm2e}

\SetAlFnt{\small}
\SetAlCapFnt{\small}
\SetAlCapNameFnt{\small}
\SetAlCapHSkip{0pt}
\IncMargin{-\parindent}
\usepackage{ textcomp }
\usepackage{ comment }
\usepackage[color=yellow!60,textsize=footnotesize,obeyDraft,draft]{todonotes}
\usepackage{colortbl}
\usepackage{booktabs}
\usepackage{capt-of}
\usepackage{makecell}
\usepackage{multirow}
\usepackage{listings}
\usepackage{graphicx}
\usepackage{amsmath}
\usepackage[font=footnotesize,labelfont=bf]{subcaption}
\usepackage[hidelinks]{hyperref}
\usepackage[capitalize]{cleveref}
\usepackage{tikz}
\usepackage{pgfplots}
\usepackage[para,online,flushleft]{threeparttable}
\usepackage{wasysym}
\usepackage{amssymb}
\usetikzlibrary{arrows}
\usepackage{breakcites}
\usepackage{array}
\usepackage{color}
\usepackage{balance} 
\usepackage{lastpage}
\usepackage{dblfloatfix}
\usepackage{authblk}
\usepackage{siunitx}
\usepackage{xurl} 
\usepackage[shortlabels]{enumitem}

\usepackage{float}
\floatstyle{plaintop}
\restylefloat{table}
\begin{document} 

\title{Integrating Formal Verification and Simulation-based Assertion Checking in a Corroborative V\&V Process}

\titlerunning{Corroborative V\&V Process}
\authorrunning{Schwammberger, Harper, Vaz Alves, Chance, et al.}

\author{Maike Schwammberger\inst{1}, Christopher Harper\inst{2}, Gleifer Vaz Alves\inst{3}, Greg Chance\inst{4}, Tony Pipe\inst{2}, Kerstin Eder\inst{4}}

\institute{University of Oldenburg, Oldenburg, Germany \and University of the West of England, Bristol, UK \and Federal University of Technology, Parana, Brazil \and University of Bristol, Bristol, UK}
\tocauthor{Authors' Instructions}

\maketitle
\let\thefootnote\relax\footnotetext{
\textit{Statements about authorship contribution.}
Maike Schwammberger (e-mail: schwammberger@informatik.uni-oldenburg.de) is with the University of Oldenburg, Oldenburg, Germany. M.\ Schwammberger was supported by the German Research Council (DFG) in the PIRE Projects SD-SSCPS and ISCE-ACPS under grant no.\ FR 2715/4-1 and FR 2715/5-1. 
Christopher Harper (e-mail: chris.harper@brl.ac.uk),
Tony Pipe (e-mail: tony.pipe@brl.ac.uk), 
are with the University of the West of England, Frenchay, Coldharbour Ln, Bristol, BS34 8QZ, United Kingdom. 
Gleifer Vaz Alves (e-mail: gleifer@utfpr.edu.br), is with the Federal University of Technology, Parana, Brazil. 
Greg Chance (e-mail: greg.chance@bristol.ac.uk), 
and 
Kerstin Eder (e-mail: kerstin.eder@bristol.ac.uk) 
are with the Trustworthy Systems Lab, Department of Computer Science, University of Bristol, Merchant Ventures Building, Woodland Road,  Bristol, BS8 1UQ, United Kingdom. G.\ Chance and K.\ Eder were supported by the ``UKRI Trustworthy Autonomous Systems Node in Functionality''  under grant number EP/V026518/1.}

\makeatletter
\renewcommand\subsubsection{\@startsection{subsubsection}{3}{\z@}%
                       {-18\p@ \@plus -4\p@ \@minus -4\p@}%
                       {4\p@ \@plus 2\p@ \@minus 2\p@}%
                       {\normalfont\normalsize\bfseries\boldmath
                        \rightskip=\z@ \@plus 8em\pretolerance=10000 }}
\makeatother

\textbf{\textit{Abstract}--Automated Vehicles (AVs) are rapidly maturing in the transportation domain. However, the complexity of the AV design problem is such that no single technique is sufficient to provide adequate validation of key properties such as safety, reliability or trustworthiness. 
In this vision paper, a combination of a spatial traffic logic and agent-based verification methods with a validation method that uses assertion checking of simulations is proposed.
We sketch how to integrate the respective approaches within a methodological framework called Corroborative Verification and Validation (V\&V).
The Corroborative V\&V framework identifies three different verification and validation levels for AVs (formal verification, simulation-based testing, real-world experiments) and specifies connections and evidence between these levels. We define specifications for the formal relationships that must be established between processes, system models and requirements models for the evidence from formal design verification and simulation-based testing to corroborate each other and enhance assurance confidence from verification and validation.}

\section{Introduction}
As a future with Automated Vehicles (AVs) comes closer each day, it is of the utmost importance to ensure their safety, reliability and trustworthiness. In this paper, we present our vision to investigate methods for combining formal verification with simulation-based testing to establish an integrated methodology for the safety validation of AVs.
We envision to apply a verification and validation philosophy called Corroborative Verification and Validation (V\&V), an approach developed previously at the Bristol Robotics Laboratory~\cite{corroborative-approach}. The Corroborative V\&V seeks to provide verification and validation assurance through the transformation and cross-checking of verification and validation evidence at different abstraction levels of design description and modelling.

\begin{figure*}[]
    \centering
    \includegraphics[
    width=0.85\linewidth
    ]{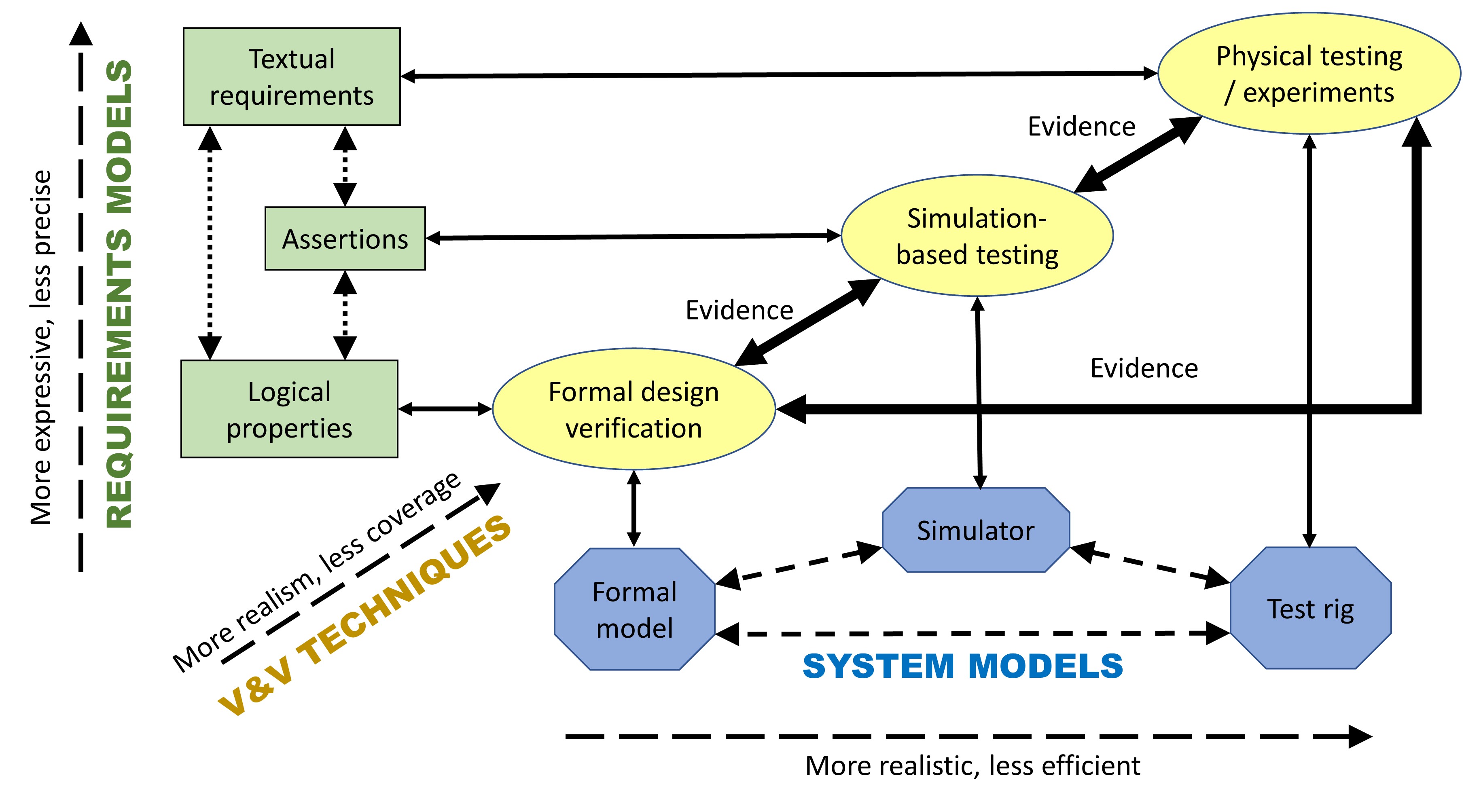}
    \caption{Lifecycle process model for Corroborative V\&V, adapted based on~\cite{corroborative-approach}.}
    \label{fig:corroborative}
\end{figure*}

Complex highly automated systems such as AVs have many safety properties that must be validated prior to their entry into service. While design verification by formal proof of properties provides high-quality assurance evidence, proofs are generally only performed for individual properties and require significant effort to perform. Formal analyses of multiple properties together can become very complex, requiring major effort. For this reason, we plan to investigate how to use individual property proofs to identify edge cases for simulation-based testing, to show that the individual properties hold as a set or to investigate system behaviour at boundary conditions or where properties may conflict with one another.

We plan to investigate the integration of formal design verification methods applied to AVs, using techniques such as Urban Multi-lane Spatial Logic (UMLSL)~\cite{Sch18-TCS}, developed at the University of Oldenburg, and agent-based verification methods~\cite{jsan10030041}, developed at Federal University of Technology -- Parana, with methods for developing assertion checks developed at the Bristol Robotics Laboratory~\cite{Harper21} for extracting simulation test cases from textual requirements such as ``the rules of the road'' documents (for example, the UK Highway Code (UKHC)~\cite{departmentfortransportusing2017}). For this, we also make use of a road traffic rule extension of UMLSL called Urban Spatial Logic for Traffic Rules (USL-TR), with which some UK Traffic Rules have been formalised \cite{schwammbergerextending2021}.

Formal verification and simulation can be thought of as complementary tools or techniques for checking system correctness with respect to a specification. Some authors have used simulators to build or determine variable values for~\cite{hoffmann2016autonomous} formal models of complex systems such as AVs. During simulation-based verification the parameter space is `sampled' but for formal verification the entire space is `proven' against a single property. Goldberg~\cite{goldberg2008bridging} attempts to integrate these techniques through use of the satisfiability problem\footnote{\url{https://en.wikipedia.org/wiki/Boolean_satisfiability_problem}} to derive a sufficient sample set that has `enough power' to prove the logical satisfiability of the simulation model with respect to a formal model. There is also empirical evidence of the corroboration of simulation and formal techniques to support the verification for a self-driving application \cite{domenici2017integrated} and in a robot handover task \cite{corroborative-approach}. Yet Chen et al.~\cite{7588753} present the integration of tools developed for railway systems, the \textit{BRaVE} tool, a railway simulator, and \textit{OnTrack}, a tool used for automatic verification of safety properties for scheme plans using formal methods. The authors intend to use a single environment for prototyping, concept, development and safety analysis. 

Our contribution is structured as follows. In Sect.~\ref{sec:corroborativevv}, we give an overview of the Corroborative V\&V philosophy and in Sect.~\ref{sec:traffic-agents}, we briefly introduce the spatial traffic rule logic USL-TR and some steps into the direction of the verification of timed-automata traffic rule controllers. We provide details on assertion checking of simulations in Sect.~\ref{sec:assertions}, followed by a sketch of our vision to combine the approaches from Sects.~\ref{sec:traffic-agents} and \ref{sec:assertions} in Sect.~\ref{sec:vision} under the roof of the Corroborative V\&V methodology from Sect.~\ref{sec:corroborativevv}.

\section{Corroborative V\&V}\label{sec:corroborativevv}

The problem of automated vehicle V\&V is complicated by the fact that validation of situated behaviour in extended domains requires a very large state space, in which all the relative states between the AV and features of the environment (other agents, static features, ambient conditions, etc.) must be defined. This leads to combinatorial expansion in the number of states to be covered, which rapidly becomes impracticable to test to acceptable coverage levels (even with automated testing). Therefore, it is critical to select specific test cases that deliver important information about the system properties of interest such as safety or reliability rather than simply performing tests mechanically across the state space.

One intention of Corroborative V\&V, as illustrated in Fig.~\ref{fig:corroborative}, is to assist the process of identifying significant test cases by using information derived from V\&V activities at different levels of design refinement to cross-check each other, and provide a means of revealing the test cases that are more significant to the validation of key system properties. The framework presented in Fig.~\ref{fig:corroborative} considers three levels of V\&V activity: static formal verification of designs by model checking or theorem proving, simulation-based testing,  and experiments with physical test rigs or road test vehicles.

\begin{figure*}[]
    \centering
    \includegraphics[width=0.85\linewidth, trim=0 0.8cm 0 2.1cm, clip]{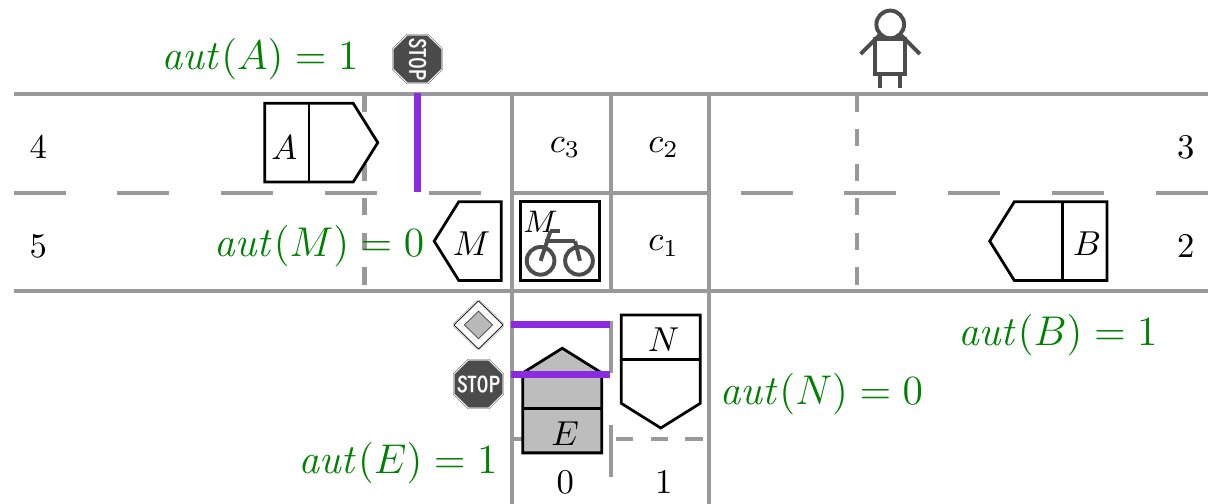}
    \caption{An example traffic situation showing traffic signs and several automated ($aut(E)=1$) and non-automated ($aut(E)=0$) road users $A$, $B$, $E$, $M$ and $N$.}
    \label{fig:traffic-situation}
\end{figure*}

The principal trade-off between these levels of activity
lies between coverage and realism. At the level of systems design, formal verification of system properties can achieve exhaustive coverage over the domains over which a design is proven. However, typically only individual properties are proven, and proofs are often grounded in assumptions or other axiomatic statements that can only be validated externally to the proof itself. At the level of the complete system, physical tests have the advantage of being realistic and therefore provide direct evidence of system safety properties. But the scope of this evidence is limited only to the system and environmental states covered in the specific test cases performed, and typically the cost and time involved in the achievement of effective coverage of all conditions are either too expensive or time-consuming to be practicable.

Simulation-based testing lies between these two levels of capability; in providing a dynamic test of a complete system model it can evaluate multiple system properties simultaneously (using assertion checking techniques~\cite{Harper21}) over any range of environmental conditions that can be simulated, and automated testing over large sets of test conditions can be performed. Such evaluation does not constitute formal proof, but specific simulation tests could reveal conditions that challenge or overturn proofs generated at the design verification stage, especially in any underlying axiomatic statements (assumptions). The principal difficulty with simulation lies in the fidelity of the simulator~\cite{determinism}; 
inaccuracies in modelling the real world may lead to mis-evaluation of system properties, possibly leading to false-positive claims that they are valid. So, the behaviour of systems in simulation must be verified against physical test data to validate the simulation.

By using each type of V\&V process judiciously, to provide cross-correction of the others, the disadvantages of any one can be compensated by the advantages of the others, and the overall quality of evidence can result in enhanced assurance of safety or any other important system property. In this paper we discuss the issues involved in integrating formal design verification with simulation-based testing; integration with physical testing is left for future work.

\section{From a Traffic Logic to the Verification of Autonomous Agents}\label{sec:traffic-agents}
We intend to build automated agents capable of knowing and recognising road traffic rules. For that, we define several steps, starting with the representation of traffic rules using a specific logic, USL-TR \cite{schwammbergerextending2021}, designed to abstract the existent elements of traffic rules as they are written in a given  Highway Code \cite{departmentfortransportusing2017}. With USL-TR, it is possible to express spatial elements of traffic rules. Such spatial elements could, e.g., be that there is a stop sign or a pedestrian crossing the road ahead.

We consider the traffic situation depicted in Fig.~\ref{fig:traffic-situation} and the AV, termed the ego car $E$, as a running example. Now take a simple fragment of the traffic rule (\textbf{170}) from the UKHC \cite{departmentfortransportusing2017}:
``\textit{Do not cross or join a road \textbf{until} there is a \textbf{gap} large enough for you to do so safely.}''
Using USL-TR, we abstract the spatial elements of this formula. For instance, a safe gap for the ego car $E$ can be represented by $\mathit{sg}(E) \,\equiv\, \mathit{free} \,\land\, \ell >= \mathit{size}_E$
, where $\mathit{free}$ indicates that there is free space on the road, and $\ell >= \mathit{size}_E$ specifies that this free space is larger than the size of $E$. Using a spatial connector $\frown$ from USL-TR, similar to the chop operator $;$ of Interval Temporal Logic \cite{Mos85}, we could then express that this free space is, e.g., in front of the ``reservation'' (the occupied space) of $E$ with a formula $re (E) \frown \mathit{sg(E)}$. In Fig.~\ref{fig:traffic-situation}, this formula would not hold, as there is not enough free space in front of $E$ due to the road user $M$ currently occupying the intersection. With an ``autonomy flag'' $\mathit{aut}(E)=1$ it is indicated that $E$ is an automated road user and $\mathit{aut}(M)=0$ states that $M$ is a non-automated road user (e.g. a cyclist).

\begin{figure*}[]
    \centering
    \includegraphics[width=0.85\linewidth]{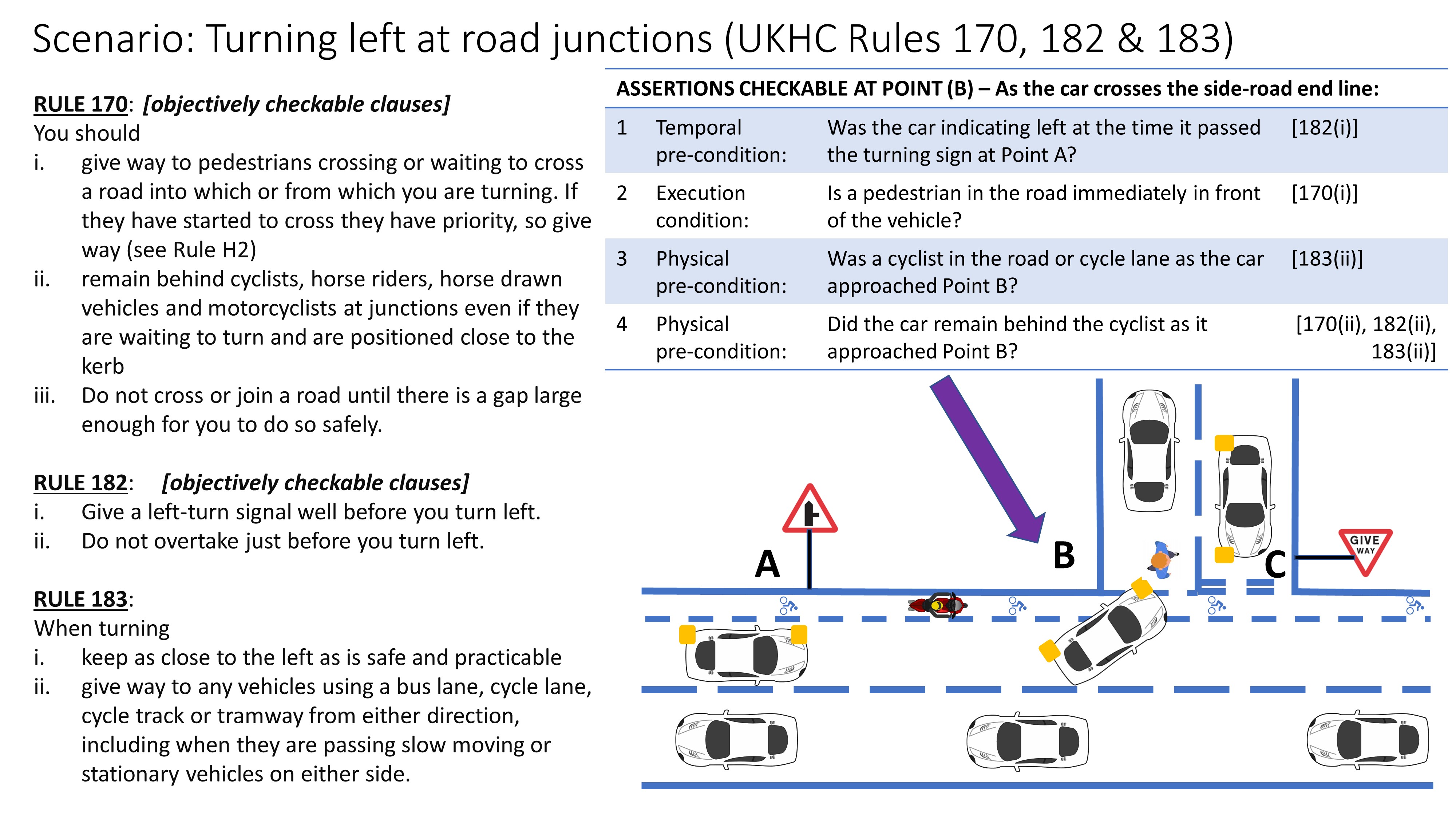}
    \caption{Example assertions based on UK Highway Code rules.}
    \label{fig:UKHC-Rule-170}
\end{figure*}

In current work~\cite{AS22}, we consider the temporal sequence of elements in a traffic rule by representing them using behaviour diagrams. At this stage, we do not yet add formal details to these diagrams but reflect the flow of actions of the rules as they are. These diagrams can be perceived as an intermediate step towards a fully formalised traffic rule. By annotating the behaviour diagrams with USL-TR formulae, we reach the first step for our planned translation from behaviour diagrams into timed automata \cite{AD94}. The following step is to add clock constraints for expressing timing behaviour contained within traffic rules (e.g. that braking or setting a turn signal are not done immediately but take some time).

Applying the previous steps, we obtain a formal, timed-automaton representation of traffic rules.
From this, we build the corresponding autonomous agents endowed with traffic rules.
We will then apply model checking techniques to these agents and formally verify properties, both for the timed automata and autonomous agents, starting from previous work in~\cite{BS19} and~\cite{jsan10030041}. Retaking the previous example, we could check properties such as ``\textit{Always when there is a stop sign, the agent first stops and proceeds with their manoeuvre when there is a safe gap large enough to do so}''. Our overall goal is to introduce a \emph{Digital Highway Code} that comprises a digitalised, precise and unambiguous version of the existing natural language road traffic regulations for AVs.

\section{Validation by Assertion Checking of Simulations}\label{sec:assertions}

In previous work \cite{Harper21} we investigated the methodology of deriving safety validation assertions from driving manuals or codes of practice, specifically the UKHC. We developed two distinct methods for constructing assertions: direct translation of natural language, and model-based analysis.

Many UKHC assertions define independently observable conditions or situations of the ego car being tested, and can be measured objectively. Assertions of this type are usually defined as conditions relative to some \emph{assertion reference point}, a time-step in the simulation at which the trigger for the rule can or should be applied. We have found that these assertions can be classified into one of four principal types: 
\begin{itemize}
    \item \emph{Invariant Condition}: A condition to be satisfied at all time steps within the captured data trace.
    \item \emph{Execution Condition}: A condition to be satisfied at the assertion reference point.
    \item \emph{Pre-condition (physical or temporal)}: A (physical or temporal) condition to be satisfied in the steps preceding the assertion reference point.
    \item \emph{Post-condition (physical or temporal)}: A (physical or temporal) condition to be satisfied in the steps following the assertion reference point.
\end{itemize}

The following example in Fig.~\ref{fig:UKHC-Rule-170} shows the identification of several assertion hypotheses derived from UKHC rule 170 and others relating to the same road environment presented in Section~\ref{sec:traffic-agents} and in Fig.~\ref{fig:traffic-situation}:

As discussed in \cite{Harper21}, the hypotheses are then transformed into queries and can be executed in a database in which the simulation data trace has been stored. The illustration shows how several different assertions may apply in any given scenario. While an automated vehicle control system design may have been verified to satisfy each rule in isolation, assertion checking in simulation can verify whether all the verified properties are satisfied together in the scenario. If an assertion is triggered at any time step, the causes of the assertion failure can be examined in the data.

\section{Vision - Corroborating Design Verification with Simulation Testing}\label{sec:vision}

Our motivation for this paper is to investigate how to integrate the approaches that have been sketched in the previous Sects.~\ref{sec:traffic-agents} and \ref{sec:assertions} using the Corroborative V\&V framework from Sect.~\ref{sec:corroborativevv}. 
The goal of the framework is to provide evidence from each of the V\&V processes, such that evidence generated by one process supports evidence generated by the others. When taken together, confidence in the overall safety (or other properties) is improved. Taking the subset of elements of Fig.~\ref{fig:corroborative} related to these two processes, and expanding the inter-relationships (arrows), we consider the following lattice diagram to capture the precise relationships between models, requirements and processes:
\begin{figure*}[]
    \centering
    \includegraphics[width=0.85\linewidth]{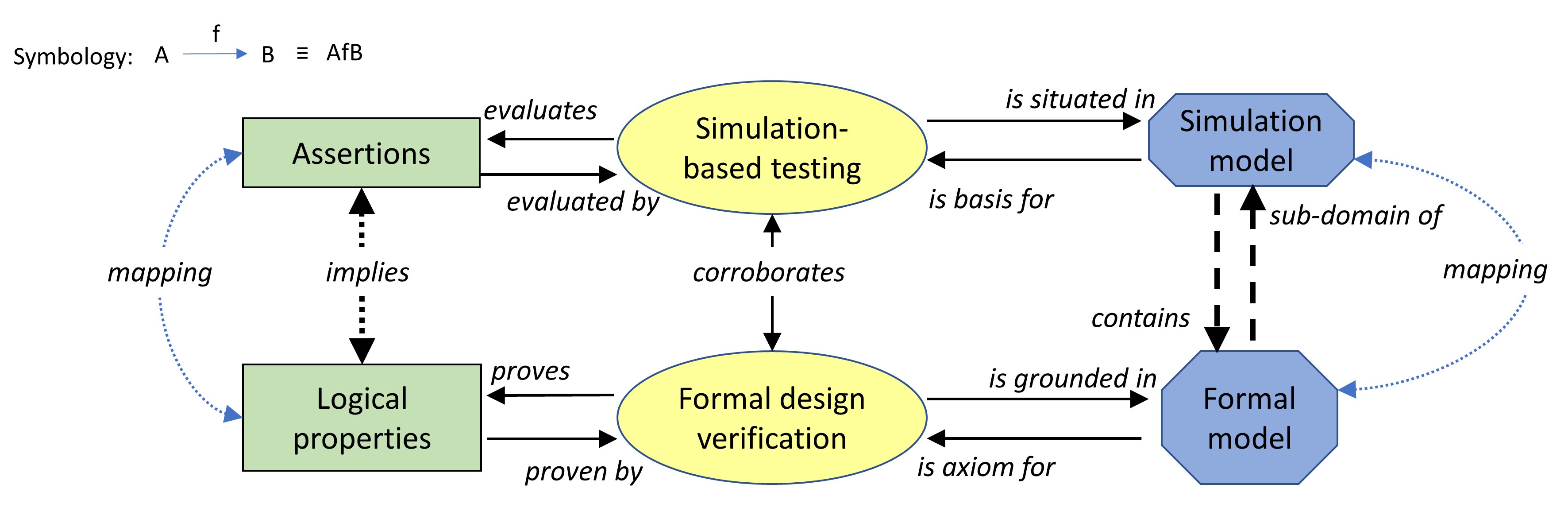}
    \caption{Relationships between Formal Design Verification and Simulation-based Testing 
    }
    \label{fig:Lattice-diagram}
\end{figure*}

Our objective is to define what is needed to ensure that evidence from Formal Design Verification corroborates that of Simulation-based Testing, and vice versa. In general parlance, the term ``corroborate" means `to provide information in support of' some statement; this can be taken to mean that if evidence from one process is found to be consistent with another, then the two corroborate one another. The corroborative V\&V framework was used in \cite{corroborative-approach} in such a manner. However, the term has been applied by Popper \cite{popper1997logic} to mean that \emph{an attempt has been made to falsify the evidence, which has failed}, thereby offering stronger support for the claims supported. This is a stronger criterion, and we propose to investigate how this may be achieved within the Corroborative V\&V Framework. Possibilities include:
\begin{itemize}
    \item Testing at boundary conditions of any assumptions defined as axioms in the formal design verification (e.g. road network, traffic conditions constraints).
    \item Testing where different (possibly conflicting) actions are defined for similar model conditions by proofs of different properties (so-called ``trolley problems'' \cite{Bonnefon1573}). 
    \item Testing at boundary conditions of initial conditions within which safety properties have been formally verified.
    \item For the planned agent model-checking (cf.\ Sect.~\ref{sec:traffic-agents}), it is of high interest to find a meaningful, maximally representative model to start with, which may be found via simulations.
    \item Other cases where different formal properties may conflict.
\end{itemize}

By combining our simulation- and verification-based approaches within the Corroborative V\&V framework via the sketched steps, our vision is that both types of approaches will be able to support (``corroborate'') each other via evidence that is exchanged between them.

\balance

\printbibliography


\end{document}